\documentclass[aps,reprint,amsmath,amssymb,superscriptaddress,showpacs,prb]{revtex4-1}
\usepackage{graphicx}
\usepackage{color}
\usepackage{lipsum}
\usepackage{dcolumn}

\begin{document}

% Title of the article
\title{Percolation description of charge transport in amorphous oxide semiconductors}

\author{A.~V.~Nenashev}
%\email {nenashev@isp.nsc.ru}
\affiliation{Institute of Semiconductor Physics, 630090 Novosibirsk, Russia}
\affiliation{Novosibirsk State University, 630090 Novosibirsk, Russia}

\author {J.~O.~Oelerich}
\affiliation{Department of Physics and Material Sciences Center,
Philipps-University, D-35032 Marburg, Germany}

\author {S.~H.~M.~Greiner}
\affiliation{Department of Physics and Material Sciences Center,
Philipps-University, D-35032 Marburg, Germany}

\author{A.~V.~Dvurechenskii}
\affiliation{Institute of Semiconductor Physics, 630090 Novosibirsk, Russia}
\affiliation{Novosibirsk State University, 630090 Novosibirsk, Russia}

\author{F.~Gebhard}
\affiliation{Department of Physics and Material Sciences Center,
Philipps-University, D-35032 Marburg, Germany}

\author {S.~D.~Baranovskii}
\email {baranovs@staff.uni-marburg.de}
\affiliation{Department of Physics and Material Sciences Center,
Philipps-University, D-35032 Marburg, Germany}

\date{\today}

\begin{abstract}
The charge transport mechanism in amorphous oxide semiconductors (AOS) is a matter of controversial debates. Most theoretical studies so far neglected the percolation nature of the phenomenon. In this article, a recipe for theoretical description of charge transport in AOSs is formulated using the percolation arguments. Comparison with the previous theoretical studies shows a superiority of the percolation approach. The results of the percolation theory are compared to experimental data obtained in various InGaZnO materials revealing parameters of the disorder potential in such AOS.

%The charge transport mechanism in amorphous oxide semiconductors (AOS)
  %is a matter of controversial debates.
  %Most theoretical studies so far neglected the percolation nature of the phenomenon.
  %Our percolation approach to the random band-edge model
  %for charge transport in AOS reproduces experimental data obtained
  %in various InGaZnO (IGZO) materials. The width of the band-edge disorder
  %and the bare electron mobility are the crucial parameters
  %for charge transport in these materials.
  \end{abstract}

%\pacs{72.80.Le,81.05.Fb,72.80.Ng,88.40.jr,72.20.Jv,73.50.Gr}
%Organic semiconductors

%conductivity of, 72.80.Le

%in materials science, 81.05.Fb

%organic photovoltaics, 88.40.jr

%Recombination
%in semiconductors, 72.20.Jv
%in thin films, 73.50.Gr

%72.80.Ng Disordered solids
%72.20.-i Conductivity phenomena in semiconductors and insulator
% Other possibilities:

\maketitle   % please do not remove

\section{Introduction}
\label{introduction}
Amorphous oxide semiconductors such as InGaZnO (IGZO)
systems are in the focus of intensive research due to
applications of these materials in thin-film transistors
for transparent and flexible flat-panel displays. Although charge transport plays
a decisive role for such applications, there is no agreement on the basic transport
mechanism. In particular, the percolation nature of charge transport
inherent for electrical conduction in disordered materials
has not been addressed properly thus far.
In the present work, we develop a concise description of charge transport
in AOS based on the percolation theory.

In solids, there are basically three distinct transport mechanisms,
and all of them were suggested as possible candidates for charge transport in AOS.
In the following, we discuss them briefly before we describe the
random band-edge model
that looks most plausible for amorphous oxide semiconductors.

\subsection{Band transport via extended states in the random barrier model}
\label{subsec:bandtransport}

\begin{figure}[ht]
\includegraphics[width=\linewidth]{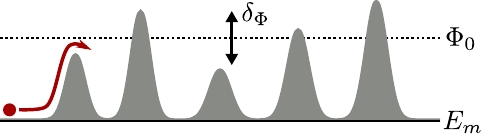}
\caption {Random barrier model for band transport above the band edge~$E_m$
  affected by random potential barriers.\label{fig:Kamiya_model}}
\end{figure}

In their pioneering works,~\cite{Nomura2004_Nature,Kamiya2009,TAKAGI2005,Kamiya2010_APL,Kimura2010_APL}
Hosono and collaborators
proposed band transport via extended states
as a possible transport mechanism in IGZO materials.
They assumed that charge carriers can move above the band edge~$E_m$
but their motion is affected by a disorder in the form of random potential barriers with a Gaussian distribution of heights,
\begin{equation}
  \label{DOS_Gauss}
  G_B(V) =  \frac{1}{\delta_{\phi}\sqrt{2\pi}}
    \exp\left(-\frac{(V-\phi_0)^{2}}{2{\delta_\phi}^{2}}\right) \; .
\end{equation}
Here, $\phi_0$ is the average height of the barriers and $\delta_{\phi}$
is the standard deviation in the distribution of the barrier heights.
This random barrier model is sketched in Fig.~\ref{fig:Kamiya_model}.

Charge transport in the random barrier model was described
in the framework of the Drude approach that is based on the average relaxation time
$\langle\tau\rangle$ for free carriers in the states above the band edge~$E_m$.
In this approximation, the carrier mobility is determined as
$\mu=e\langle\tau\rangle/m$, where $m$ is the effective mass.
For the band transport in the absence of a disorder potential,
the Drude approach leads to the expression~\cite{Sze1981}
\begin{equation}
  \label{eq:sze}
  \mu = -\frac{e}{m\cdot n}\int_{E_m}^{\infty}
  \tau(E) v_z(E) \frac{\partial{f_e(E)}}{\partial{v_z}}D_m(E){\rm d}E
   \; ,
\end{equation}
where $n$ is the concentration of carriers, $\tau(E)$ is the momentum relaxation time
at the electron energy $E$, $v_z(E)$ is the electron velocity along
the transport direction ($z$ axis), $D_m(E)$ is the density of states above the band
edge $E_m$, and $f_e(E)$ is the Fermi function.

In the presence of substantial disorder in AOS, the Drude approach
is then modified heuristically by the introduction
of the weight function~\cite{Kamiya2010_APL}
\begin{equation}
  \label{eq:Kamiya ro}
  \varrho(E) = \int_{E}^{\infty}G_B(\varepsilon){\rm d}\varepsilon
   \; ,
\end{equation}
which was termed ``transmission probability''.\cite{Kamiya2010_APL}
The expression for the charge carrier mobility in AOS thus attains the form
\begin{equation}
  \label{eq:Kamiya}
  \mu = -\frac{e}{m\cdot n}\int_{E_m}^{\infty}\tau(E) v_z(E) \varrho(E)
  \frac{\partial{f_e(E)}}{\partial{v_z}}D_m(E){\rm d}E \; .
\end{equation}
The introduction of the weight function~$\varrho(E)$ into Eq.~(\ref{eq:Kamiya})
was interpreted~\cite{Kamiya2010_APL} as taking into account
the percolation arguments suggested by Adler \textit{et al.}~\cite{Adler1973}.
%;
However, percolation has little to do  with Eqs.~(\ref{eq:Kamiya ro})
and~(\ref{eq:Kamiya}), as is readily seen from the fact that the percolation threshold
does not appear in the above equations. This feature will be discussed in Sec.~\ref{sec:percolation theory}.

More importantly, if disorder creates potential barriers above the band edge
as sketched in Fig.~\ref{fig:Kamiya_model}, it will also create potential wells
below the band edge. The statistical distribution of these wells must be taken
into account as well, which makes $E_m$ a regional, random quantity
(random band-edge model).
A description of charge transport based on percolation theory for
the random band-edge model
will be given in Sec.~\ref{sec:percolation theory}.

\subsection{Trap-limited band transport}

Trap-limited transport in the spirit of the multiple--trapping (MT) model,
has also been considered as a possible transport mechanism in AOS \cite{Park2010,LeeParkKim2010}. In
the MT process, the motion of charge carriers via delocalized states is interrupted by trapping into the localized states
with subsequent activation of carriers back into the conducting states
above the mobility edge.

The energy spectrum of localized states in the band tails of inorganic
amorphous semiconductors is widely considered to exhibit
a purely exponential
shape,\cite{Mott1979,Thomas1989,Street1991,Baranovski2006,Smeniuk2017}
\begin{equation}
\label{DOS_exp}
g(E) = N_m \exp\left(\frac{E}{E_{0}}\right) \; ,
\end{equation}
where $E$ is the energy of the trap counted from the band edge $E_m$,
$E_{0}$ is the energy scale, and $N_m$ is the density of localized states
at the band edge $E_m$.
%The value of $E_{0}$ depends on the material and it is usually
%estimated to lie below
%$0.1$ eV.\cite{Mott1979,Thomas1989,Street1991,Baranovski2006,Smeniuk2017}

Lee \textit{et al.}~\cite{LeeParkKim2010} used an extraction technique
to determine the subgap density of states in an $n$-channel amorphous
InGaZnO (a-IGZO) thin-film transistor based on the study of the multifrequency
capacitance-voltage (C--V) characteristics. They concluded that
the subgap density of states
is a superposition of two distinct coexisting exponential functions.
Lee \textit{et al.}\cite{Lee2011_APL,Lee2012_APL}
combined the above models of band transport limited by potential
barriers~\cite{Nomura2004_Nature,TAKAGI2005,Kamiya2010_APL,Kimura2010_APL}
and the multiple-trapping transport.\cite{Park2010,LeeParkKim2010}
Consequently, they assume two distinct transport regimes
in a thin-film-transistor based on AOS, depending on the
concentration of carrier.
At low gate voltages, i.e.,
at small carrier concentrations $n$, the Fermi level lies
in the manifold of the localized states, characterized by the density of states given by
Eq.~(\ref{DOS_exp}), with energies below the band edge $E_m$.
In this MT regime, the drift mobility of carriers is determined as
\begin{equation}
\label{eq:mob MT}
\mu \simeq \mu_{mod} \frac{n_{\textit{free}}}{n_{\textit{free}} + n_{\textit{ftrap}}}\; ,
\end{equation}
where $n_{\textit{free}}$ and $n_{\textit{trap}}$
are the free (above $E_m$) and trapped (below $E_m$) carrier densities,
respectively, and $\mu_{mod}$ is treated as the usual band mobility $\mu_0$,
``modulated by the percolation term''.\cite{Lee2011_APL}
Lee \textit{et al.}~\cite{Lee2011_APL} mention that this term
should be determined by the ratio between
the potential barrier height and the average barrier width.
Nevertheless, they use the relation between $\mu_{mod}$ and $\mu_0$,
derived by the averaging of transition rates for overcoming potential barriers. The latter method has been, however, qualified as not appropriate for description of incoherent charge transport \cite{Shklovskii1984,Baranovski2006}.

\subsection{Hopping transport}

The incoherent tunneling of charge carriers between localized states,
distributed randomly in space and energy, also has been suggested
as a possible charge transport mechanism in IGZO materials. \cite{Germs2012}
A marginal admixture of band transport was also assumed
in order to account for the Hall measurements.
The theoretical analysis by Germs \textit{et al.}~\cite{Germs2012}
is based on the concept of the transport energy~$E_t$.
According to this approach, charge transport in a system with localized electron states
is due to the activation of carriers from the Fermi level $E_f$ towards
the vicinity of the transport energy.\cite{Gruenewald1979,Monroe1985,Baranovskii1995,Baranovskii1997,Baranovskii2000}
The carrier mobility in this transport regime can be written
as~\cite{Baranovskii2002a,Baranovski2006,Baranovskii2014,Nenashev_Topical_2015}
\begin{equation}
\label{eq:mob_Germs}
\mu = \widetilde{\mu_0} \exp \left[-\frac{E_t - E_f(n,T)}{kT}\right] \; ,
\end{equation}
where $E_t$ is the transport energy and the
prefactor $\widetilde{\mu_0}$ depends on the concentration of carriers $n$.

To fit the high values of the carrier mobility measured in IGZO materials,
an unusually large value of the localization length in the tail states,
$a \simeq 4.8\, \text{nm}$, is needed in the model
of hopping transport.\cite{Germs2012}
This value exceeds by far the estimates for the localization length
of carriers in the band tails of inorganic
semiconductors.\cite{Mott1979,Thomas1989,Street1991,Baranovski2006}
Therefore, it appears unlikely that hopping transport is the dominant mechanism
for charge transport in AOS.
%However, the study by Germs \textit{et al.}~\cite{Germs2012}
%is the only one in the literature on charge transport in IGZO materials
%that is based on percolation theory because
%the calculation of the transport energy $E_t$ for hopping transport
%relies on percolation
%arguments.~\cite{Gruenewald1979,Rubel2004,Baranovskii2014,Oelerich2014,Nenashev_Topical_2015}

\subsection{Random band-edge model}

\begin{figure}[ht]
\includegraphics[width=\linewidth]{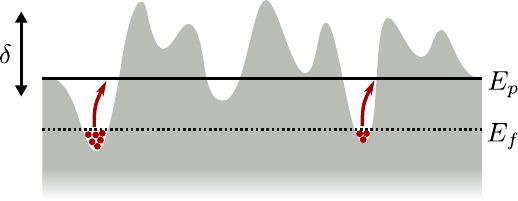}
\caption{Schematic representation of the spatial fluctuations of the band edge $E_m$
  in the random band-edge model.
  The carrier motion is due to activation from the Fermi level $E_f$ towards the
  percolation level $E_p$.\label{fig:our_Model}}
\end{figure}

Recently, Fishchuk {\it et al.}~\cite{Fishchuk2016} addressed a model
that combines band transport and localized band-tail states,
though with a significant modification.
While Kamiya \textit{et al.}~\cite{Kamiya2009,Kamiya2010_APL}
and Lee \textit{et al.}~\cite{Lee2011_APL,Lee2012_APL} assume
the distribution of potential barriers of the form given by Eq.~(\ref{DOS_Gauss})
above a global band edge $E_m$, see Fig.~\ref{fig:Kamiya_model},
Fishchuk {\it et al.}~\cite{Fishchuk2016} assume that the disorder potential
causes random long-range variations of the band edge $E_m$,
as illustrated in Fig.~\ref{fig:our_Model}.
The spatial fluctuations of the band edge $E_m$
are assumed to be Gaussian with the distribution function~\cite{Fishchuk2016}
\begin{equation}
  \label{eq:DOS_Gauss_E_m}
  G(E_m) =
  \frac{1}{\delta\sqrt{2\pi}}
    \exp \left[-\frac{1}{2}\left(\frac{E_m}{\delta}\right)^2\right] \, ,
\end{equation}
where $\delta$ is the standard deviation, and the position of the band edge $E_m$
is counted from the position of the band edge without disorder potential.
In this work,
we consider the random band-edge model
suggested by Fishchuk \textit{et al.}~\cite{Fishchuk2016}
as appropriate for the description of charge transport in AOS.

Fishchuk \textit{et al.}~\cite{Fishchuk2016} applied the
effective-medium-ap\-proxi\-mation (EMA)
to study theoretically charge transport.
In contrast, we use percolation arguments to develop a theoretical description
of charge transport in the random band-edge model.
At $kT \ll \delta$, where $kT$ is the thermal energy and $\delta$
is the scale of disorder in Eq.~(\ref{eq:DOS_Gauss_E_m}),
the results of our percolation theory are reliable and they substantially differ from those of the EMA approach used
previously \cite{Fishchuk2016}, as shown in Sec.~\ref{sec:comparison_EMA}. Therefore, 
the percolation theory seems superior to the EMA
for the description of charge transport within the random band-edge model.

\subsection{Outline}

In Sec.~\ref{sec:model}, we describe in more detail the random band-edge model of
Fishchuk \textit{et al.}~\cite{Fishchuk2016} that we employ
for our study.
In Sec.~\ref{sec:percolation theory}, we show how to calculate the carrier mobility
using percolation theory.
In Sec.~\ref{sec:comparison_EMA}, we compare our percolation approach
with the effective medium approximation.
In Sec.~\ref{sec:results}, we compare our theoretical results
with experimental data. It is shown that percolation theory
is capable to account for the dependencies of the charge carrier mobility
on temperature and on the concentration of charge carriers in IGZO materials.
Moreover, a comparison between the results of percolation theory
with experimental data reveals the characteristic parameter~$\delta$
of the band-edge disorder
in Eq.~(\ref{eq:DOS_Gauss_E_m}) and the conduction-electron mobility~$\mu_0$.

\section{Random band-edge model for charge transport in AOS}
\label{sec:model}

The random band-edge model by Fishchuk \textit{et al.}~\cite{Fishchuk2016}
assumes that the position of the band edge $E_m$ varies
in space due to disorder potential.
The distribution of $E_m$ values that belong to different spatial regions
is characterized by the Gaussian distribution $G(E_m)$ given by
Eq.~(\ref{eq:DOS_Gauss_E_m}). This distribution in the regional positions of the band
edge
plays a crucial role in the rest of this paper.

For the sake of completeness, we also include
localized states with energies below $E_m$,
whose density of states is assumed to be exponential, see Eq.~(\ref{DOS_exp}),
where the energy $E$ is counted from the band edge $E_m$.
Following Fishchuk \textit{et al.}~\cite{Fishchuk2016}, we assume that
the delocalized states with energies above $E_m$
are characterized by the density of states
\begin{equation}
  \label{eq:DOS_sqrt}
  g(E - E_m) = g_c \sqrt{E - E_m + \Delta E} \; ,
\end{equation}
where the value $g_c = 1.4 \cdot 10^{21} \, \text{cm}^{-3}\text{eV}^{-3/2}$
has been reported for a-IGZO thin films.\cite{Kamiya2010_APL}
Equations~(\ref{eq:DOS_sqrt}) and~(\ref{DOS_exp}) can be combined
to the \emph{regional\/} density of states
\begin{eqnarray}
  \label{eq:DOS_full}
  g(E - E_m) &=&
  \Theta(E-E_m) \, g_c \, \sqrt{E - E_m + (N_m/g_c)^2} \nonumber \\
  && + \big[1 - \Theta(E-E_m)\big] \, N_m \, \exp\left( \frac{E-E_m}{E_0} \right)  \; ,
  \nonumber \\
\end{eqnarray}
where $\Theta(x)$ is the Heaviside step function.
The first term on the right-hand side describes the density
of delocalized states above $E_m$,
and the second term describes the density of localized states below $E_m$.
The energy shift $\Delta E$ in Eq.~(\ref{eq:DOS_sqrt})
guarantees the continuity of the density of states at $E=E_m$
when we choose $\Delta E = (N_m/g_c)^2$.

For a given value of $E_m$, one can find a corresponding \emph{regional}
electron density,
\begin{equation}
  \label{eq:n_local}
  n_{\text{local}}(E_m) = \int_{-\infty}^{+\infty} g(E - E_m) \, f(E) \, {\rm d}E \; ,
\end{equation}
where $f(E)$ is the Fermi function,
\begin{equation}
  \label{eq:Fermi_function}
  f(E) = \left[ \exp\left( \frac{E-E_f}{kT} \right) + 1 \right]^{-1} \; ,
\end{equation}
and $E_f$ is the Fermi level. The \emph{total} electron density $n$,
averaged over the regional positions of the mobility edge $E_m$, is
\begin{equation}
  \label{eq:n_total}
  n = \int_{-\infty}^{+\infty} G(E_m) \, n_{\text{local}}(E_m) \, {\rm d}E_m \; .
\end{equation}
For given temperature $T$ and electron concentration~$n$,
the position of the Fermi level $E_f$ can be found from the system of
equations (\ref{eq:DOS_full})--(\ref{eq:n_total}).

The AOS material is assumed to be a medium
with a smoothly varying regional conductivity $\sigma_{\text{region}}$,
which is a product of the elementary charge $e$, the conduction-band electron mobility $\mu_0$,
and the local concentration of mobile electrons (with energies above $E_m$),
\begin{equation}
  \label{eq:sigma_local}
  \sigma_{\text{region}}(E_m) = e \mu_0 \int_{E_m}^{+\infty} g(E - E_m) \, f(E) \,
        {\rm d}E \; .
\end{equation}
This assumption is justified when the electron mean-free path
is small compared to the spatial scale for variations of $E_m$.

The global (macroscopic) conductivity~$\sigma$ is to be found by some ``averaging''
of the regional values $\sigma_{\text{region}}(E_m)$,
taking into account the Gaussian distribution $G(E_m)$ of the mobility edge $E_m$,
i.e., $\sigma=\langle \sigma_{\text{region}}\rangle$.
When the global $\sigma$ is found, one can calculate the (measured)
mobility $\mu$ as
\begin{equation}
  \label{eq:mu_def}
  \mu = \frac{\sigma}{en} \; .
\end{equation}
In the case of an exponentially broad scatter of regional conductivities,
a proper choice of the averaging procedure
is crucial for a correct determination of the global conductivity~$\sigma$.
We will consider three methods of ``averaging'':
the first is based on the effective medium approximation,
expressed by Eq.~(\ref{eq:effective_medium_equation_for_sigma}),
see Sec.~\ref{sec:comparison_EMA},
the other two procedures are based on percolation theory expressed
by Eqs.~(\ref{eq:sigma_percolation1}) and~(\ref{eq:sigma_Adler}),
see Sec.~\ref{sec:percolation theory}.

For the experimentally accessible regions of the $(n,T)$ phase diagram
of IGZO materials, Fishchuk \textit{et al.}~\cite{Fishchuk2016} have shown that
the conductivity $\sigma(n,T)$ and the mobility
$\mu(n,T)$ depend on
the carrier-concentration~$n$
and the temperature~$T$
mostly through the
variations of the conduction-band edge $E_m$,
and are not limited by the localized states.
Therefore, localized states might be disregarded in IGZO films.

\section{Percolation theory for charge transport in the random band-edge model}
\label{sec:percolation theory}

\subsection{From regional to global conductivities in continuum percolation theory}
\label{subsec:PCTSergei}

The random band-edge model
belongs to the class of
\textit{continuum percolation problems}.\cite{Shklovskii1984}
The transport is determined by charge carriers with energies above
the percolation level $E_p$, which is defined as the minimal energy
that allows a transport path via connected regions with $E_m$ not exceeding $E_p$.

Let $p(E)$ denote the volume fraction of regions
where the mobility edge $E_m$ is below $E$,
\begin{equation}
  \label{eq:volume_Adler}
  p(E) =  \int_{-\infty}^{E} G(E_m) \, {\rm d}E_m \; .
\end{equation}
The quantity
\begin{equation}
  \label{eq:perc_criterium}
  \vartheta_c = p(E_p)
\end{equation}
plays the role of the dimensionless percolation threshold determined
as the minimal volume fraction of the conducting material
that enables electrical connection throughout the infinitely large sample.
Numerical studies for Gaussian energy distributions
with various spatial correlation properties yield
the value $\vartheta_c=0.17 \pm 0.01$
for the three-dimensional continuum percolation
problem.\cite{Shklovskii1984}
Using the value $\vartheta_c=0.17$ in Eq.~(\ref{eq:perc_criterium}),
one obtains the value of the percolation level~$E_p$
\begin{equation}
  \label{eq:E_p}
  E_p = - 0.95 \, \delta \; .
\end{equation}
It remains to determine
the macroscopic conductivity $\sigma$ from the regional
conductivities $\sigma_{\text{region}}(E_m)$.

According to the percolation approach, regions with $E_m<E_p$
can be considered as isolated islands that are not connected with each other and,
therefore, do not contribute to $\sigma$.
The global conductivity $\sigma$ depends only on those
$\sigma_{\text{region}}(E_m)$ that obey $E_m>E_p$.
The simplest recipe to calculate $\sigma$ on the basis of percolation theory
is to average the regional conductivities
over the regions where $E_m>E_p$,
\begin{equation}
  \label{eq:sigma_percolation1}
  \sigma = \frac{1}{1-\vartheta_c} \int_{E_p}^{+\infty}
  \sigma_{\text{region}}(E_m) \, G(E_m) \, {\rm d}E_m \; ,
\end{equation}
where $G(E_m)$ is the Gaussian distribution of the local mobility edges $E_m$.
The mobility $\mu=\sigma/en$ that corresponds to
Eq.~(\ref{eq:sigma_percolation1}) can be expressed as
\begin{equation}
  \label{eq:mu_percolation1}
  \mu = \mu_0 \, \frac{n_{\text{mob}}}{n} \; ,
\end{equation}
where $n_{\text{mob}}$ is the average concentration of \emph{mobile\/}
electrons in the regions with $E_m>E_p$, and $n$ is the total electron concentration.
It is easy to recognize that eq.~(\ref{eq:sigma_percolation1})
gives the correct value of the conductivity $\sigma = e \mu_0 n$
in the absence of disorder, $\delta = 0$.

Eq.~(\ref{eq:sigma_percolation1})
also gives the correct value in the opposite limit of very pronounced disorder,
$kT \ll \delta $, for a non-degenerate occupation of states above $E_p$,
when the regional conductivities $\sigma_{\text{region}}(E_m)$
have an exponentially broad distribution of values.
In this case, the Fermi function can be approximated
as $f(E)=\exp\big[(E_f-E)/kT\big]$ and Eq.~(\ref{eq:sigma_local})
yields the exponential dependence
\begin{equation}
  \label{eq:sigma_local_nondegenerate}
  \sigma_{\text{region}}(E) = e \mu_0 N_c \exp\left( \frac{E_f-E}{kT} \right) \; ,
\end{equation}
where $N_c$ is the effective density of states in the conduction band,
\begin{equation}
  \label{eq:Nc}
  N_c = \int_0^{+\infty} g(E) \exp(-E/kT) \, {\rm d}E \; .
\end{equation}
Inserting Eq.~(\ref{eq:sigma_local_nondegenerate})
into Eq.~(\ref{eq:sigma_percolation1}) gives the asymptotic expression
for the carrier mobility
\begin{equation}
\label{eq:mu_percolation1_nondegenerate}
\mu = \frac{\sigma}{en} = \frac{\mu_0 N_c}{n(1-\vartheta_c)}
\int_{E_p}^{+\infty} \exp\left( \frac{E_f-E}{kT} \right) G(E) \, {\rm d}E \; .
\end{equation}
At low temperatures, $kT \ll \delta$,
the main contribution to the integral comes from the vicinity of the
percolation level $E_p$.
Therefore, one can approximately replace $G(E)$ by $G(E_p)$
and take the constant factor $G(E_p)$ out of the integral.
The remaining integral is elementary and
the carrier mobility given by Eq.~(\ref{eq:mu_percolation1}) assumes its
asymptotic form
\begin{equation}
  \label{eq:mu_percolation1_assymptot}
  \mu \approx \mu_0 \frac{N_c}{n} \frac{G(E_p) \, kT}{1-\vartheta_c}
  \exp\left( \frac{E_f-E_p}{kT} \right) \; .
\end{equation}
The exponential term in Eq.~(\ref{eq:mu_percolation1_assymptot})
shows that the charge transport is dominated by thermal activation of electrons
to the percolation level $E_p$,
as schematically depicted in Fig.~\ref{fig:our_Model}.
More sophisticated considerations~\cite{Shklovskii1984,Nenashev2013}
lead to a marginal correction of the pre-exponential factor in this equation.
In fact, the pre-exponential factor should contain $(kT)^\nu$ instead of $kT$
where $\nu \simeq 0.88$ is the critical exponent
for the correlation length of the percolation
cluster.\cite{Wang2013,Hao2014,Koza_2016}
Below, we will use Eq.~(\ref{eq:mu_percolation1_assymptot})
and ignore this marginal correction.

\subsection{Averaging procedure by Adler et al.}

In several theoretical studies of charge transport in AOS,
the percolation approach suggested by Adler \textit{et al.}\cite{Adler1973}
has been invoked.\cite{Kamiya2010_APL,Germs2011}
Adler \textit{et al.}~\cite{Adler1973} considered a system with a
Gaussian distribution of the regional band edges as given
by Eq.~(\ref{eq:DOS_Gauss_E_m}). They
suggested that the global conductivity $\sigma$ can be obtained as
\begin{equation}
  \label{eq:sigma_Adler}
  \sigma_{\text{A}} =
  \frac{1}{kT} \int_{E_p}^{+\infty}\sigma_{\text{A}}(E) f(E) \big[1-f(E)\big] \, {\rm d}E \;.
\end{equation}
Here $\sigma_{\text{A}}(E)$
is the contribution to the conductivity of carriers with energy $E$,
\begin{equation}
  \label{eq:sigma_E_Adler}
  \sigma_{\text{A}}(E) = \tilde B \big[p(E)-\vartheta_c\big]^2 ,
\end{equation}
where $\tilde B$ is some unspecified constant.

For a comparison with our approach in Sec.~\ref{subsec:PCTSergei},
we analyze the conductivity $\sigma_{\text{A}}$ from
Eq.~(\ref{eq:sigma_Adler})
when the Fermi level is far below the percolation level, $E_f < E_p$ and $kT \ll E_p-E_f$.
Then, we can use the Boltzmann approximation
$f(E) \approx \exp\big[(E_f-E)/kT\big]$ and $1-f(E) \approx 1$ for the Fermi functions
in Eqs.~(\ref{eq:sigma_Adler}) and~(\ref{eq:sigma_E_Adler})
in the non-degenerate case, leading to
\begin{equation}
  \label{eq:sigma_Adler_nondegenerate2}
  \sigma_{\text{A}} = \frac{\tilde B}{kT}  \int_{E_p}^{+\infty}
  \exp\left(\frac{E_f-E}{kT}\right) \big[p(E)-\vartheta_c\big]^2 \, {\rm d}E \; .
\end{equation}
At low temperatures, $kT \ll \delta$,
the major contribution to this integral comes from the region $E \approx E_p$,
providing $G(E) \approx G(E_p)$.
Using Eqs.~(\ref{eq:volume_Adler}) and~(\ref{eq:perc_criterium}),
the factor $\big[p(E)-\vartheta_c\big]$ can be simplified to
\begin{equation}
\label{eq:p_minus_theta_c}
  p(E)-\vartheta_c =  \int_{E_p}^{E} G(E') \, {\rm d}E'  \approx  (E-E_p) G(E_p) \; .
\end{equation}
Concomitantly, Eq.~(\ref{eq:sigma_Adler_nondegenerate2}) simplifies to
\begin{equation}
  \label{eq:mu_percolation2_nondegenerate_}
  \sigma_{\text{A}}
  \approx \frac{\tilde B}{kT} \big[G(E_p)\big]^2 \int_{E_p}^{+\infty}
  \exp\left( \frac{E_f-E}{kT} \right) \, (E-E_p)^2 \, {\rm d}E \; .
\end{equation}
The integral is elementary.
Using its value in Eq.~(\ref{eq:mu_percolation2_nondegenerate_}),
we obtain the asymptotic expression for the mobility
\begin{equation}
  \label{eq:mu_percolation2_assymptot}
  \mu_{\text{A}} \approx  \frac{2 \tilde B [G(E_p)]^2 (kT)^2}{en}
  \exp\left( \frac{E_f-E_p}{kT} \right) \; .
\end{equation}
A comparison of this expression with the result of percolation theory in
Eq.~(\ref{eq:mu_percolation1_assymptot}) shows that
the exponential term is correctly reproduced by
Eq.~(\ref{eq:mu_percolation2_assymptot}). However,
besides the unknown coefficient $\tilde B$,
Eq.~(\ref{eq:mu_percolation2_assymptot}) displays
an incorrect temperature dependence of the pre-exponential factor
due to the assumption of a quadratic energy dependence of the regional conductivity
above the threshold, see Eq.~(\ref{eq:sigma_E_Adler}).

%The exponential term $\exp[ (E_f-E_p)/(kT)]$
%in Eq.~(\ref{eq:mu_percolation2_assymptot}),
%and the dependence of the mobility in Eqs.~(\ref{eq:Kamiya ro}) and~(\ref{eq:Kamiya})
%on $\exp[(E_f-E_m)/(kT)]$
%suggest that the approach by Adler \textit{et al.}~\cite{Adler1973}
%may have been misinterpreted. 
The (global) band-edge $E_m$
in the random-barrier model, as
described in Sec.~\ref{subsec:bandtransport},
and the percolation threshold $E_p$ are unrelated conceptually.
However, the question remains:
is it possible to interpret the energy level $E_m$
in the model sketched in Fig.~\ref{fig:Kamiya_model}
as the percolation level $E_p$ in Fig.~\ref{fig:our_Model}
so that the random-barrier model can be viewed as part of the
random band-edge model for energies above the percolation level, $E>E_p$?
Unfortunately, this is not the case. First,
the random-barrier model in Fig.~\ref{fig:Kamiya_model} cannot contain a recipe
on how to calculate the percolation level $E_p$.
Second, 
%the energies $E$ in Eqs.~(\ref{eq:sze}) and~(\ref{eq:Kamiya})
%should be counted from the position of the band edge $E_m$.
since $E_m$ is a regional feature determined by the
distribution~(\ref{eq:DOS_Gauss_E_m}), one cannot consider the
value $E$ in Eqs.~(\ref{eq:sze}) and~(\ref{eq:Kamiya}) as if $E_m$
were uniform for the whole system. Therefore, an approach based on replacing $E_m$
in Fig.~\ref{fig:Kamiya_model} by the percolation level $E_p$
would not make sense.

In Sec.~\ref{sec:results}, we will compare
the predictions of the percolation theory expressed by
Eq.~(\ref{eq:sigma_percolation1}) with experimental data
obtained by several experimental groups
on the dependences of the carrier mobility $\mu(n,T)$
on the carrier concentration~$n$
and temperature $T$
in IGZO materials.\cite{Kamiya2009,Kamiya2010_APL,Germs2012,Fishchuk2016}.
Before that
we address in the next Sec.~\ref{sec:comparison_EMA} the relation
between the results from percolation theory and those
based on the effective-medium-approximation (EMA).

\section{Comparison between percolation theory and EMA}
\label{sec:comparison_EMA}

%In their work, Fishchuk \textit{et al.}~\cite{Fishchuk2016}
%pointed to percolation theory as a suitable tool to describe charge transport
%in strongly inhomogeneous disordered systems.
%Assuming a case of moderate disorder, they applied the effective medium approximation (EMA)
%to average the regional conductivities for their study
%of charge transport in the IGZO materials
%in the framework of the random band-edge model.

The EMA and percolation theory are often considered
as complementary to each other in their ability
to account for charge transport in disordered systems.
Percolation theory is considered to be valid for strongly disordered
systems,\cite{Shklovskii1984} while in systems with a weak disorder
the EMA is often applied.\cite{Fishchuk2016}
In fact, percolation theory gives reliable results
not only for the case of strong disorder, $kT\ll\delta$,
but also for the opposite case of $\delta \rightarrow 0$,
as discussed in Sec.~\ref{sec:percolation theory}
in the context of Eq.~(\ref{eq:sigma_percolation1}).
Therefore, it is instructive to estimate the difference between the results
of percolation theory and those of the EMA in the case of strong disorder.

In the EMA framework used by Fishchuk \textit{et al.}~\cite{Fishchuk2016}
the conductivity $\sigma$ was determined
from its regional values $\sigma_{\text{region}}(E_m)$ via the equation
\begin{equation}
  \label{eq:effective_medium_equation_for_sigma}
  \left\langle  \frac{\sigma_{\text{local}}-\sigma}{\sigma_{\text{local}}+(d-1)\sigma}
  \right\rangle = 0 \; ,
\end{equation}
where $d$ is the spatial dimension,
and the angular brackets mean the averaging
over the density distribution function $G(E_m)$
\begin{equation}
  \label{eq:meaning_of_brackets}
  \left\langle  \mathcal{A}  \right\rangle  \equiv  \int_{-\infty}^{+\infty}
  G(E_m) \mathcal{A}(E_m) \, {\rm d}E_m \; .
\end{equation}
To calculate the carrier mobility $\mu(n,T)$ in the framework of the EMA,
one should calculate the Fermi level $E_f$ from
Eqs.~(\ref{eq:DOS_full})--(\ref{eq:n_total}), and
then determine the dependence of the regional conductivity
$\sigma_{\text{region}}(E_m)$ on the regional mobility edge $E_m$,
which then leads to the global conductivity $\sigma$
via Eq.~(\ref{eq:effective_medium_equation_for_sigma}) and to the
carrier mobility $\mu$ via Eq.~(\ref{eq:mu_def}).

Let us rewrite the averaging condition~(\ref{eq:effective_medium_equation_for_sigma})
in the following equivalent form in three dimensions, $d=3$,
\begin{equation}
  \label{eq:effective_medium_X_def}
  \left\langle  \frac{\sigma_{\text{region}}}{\sigma_{\text{region}}+2\sigma}
  \right\rangle =  \frac{1}{3} \; .
\end{equation}
In the limit of low temperature and low electron concentration,
the dependence $\sigma_{\text{region}}(E_m)$
of the regional conductivity on the regional conduction band edge $E_m$
is very steep, i.e.,
for almost all values of $E_m$ we have
either $\sigma_{\text{region}}(E_m) \gg 2\sigma$ or
$\sigma_{\text{region}}(E_m) \ll 2\sigma$.
In the first case, the expression inside the angular
brackets in Eq.~(\ref{eq:effective_medium_X_def}) is close to unity,
in the second case, it is close to zero. Therefore,
\begin{equation}
  \label{eq:effective_medium_sigma_local_approx}
  \frac{\sigma_{\text{local}}(E_m)}{\sigma_{\text{local}}(E_m)+2\sigma}  \approx
  \begin{cases}
   1 & \text{if } E_m < E^* \; , \\
   0 & \text{if } E_m > E^* \; ,
  \end{cases}
\end{equation}
where $E^*$ is the value of $E_m$ that separates these two limits,
\begin{equation}
  \label{eq:sigma_local_E_star_def}
  \sigma_{\text{region}}(E^*) = 2\sigma \; .
\end{equation}
Inserting Eq.~(\ref{eq:effective_medium_sigma_local_approx})
into Eq.~(\ref{eq:effective_medium_X_def}),
and using the rule  of averaging~(\ref{eq:meaning_of_brackets}),
one can evaluate Eq.~(\ref{eq:effective_medium_X_def}) as
\begin{equation}
  \label{eq:sigma_local_X_via_E_star}
  \int_{-\infty}^{E^*} G(E_m) \, {\rm d}E_m = \frac{1}{3} \; .
\end{equation}
This equation defines the energy $E^*$.

In the case of Gaussian distribution function $G(E_m)$,
Eq.~(\ref{eq:DOS_Gauss_E_m}), the solution is
\begin{equation}
  \label{eq:E_star_value}
  E^*  \approx  -0.43 \, \delta \; .
\end{equation}
With this value for $E^*$, one can find the macroscopic
conductivity $\sigma$ from Eq.~(\ref{eq:sigma_local_E_star_def}),
where $\sigma_{\text{region}}(E^*)$ is to be calculated from
Eq.~(\ref{eq:sigma_local}).
Inserting Eq.~(\ref{eq:sigma_local_nondegenerate})
into Eq.~(\ref{eq:sigma_local_E_star_def})
provides the following asymptotic expression for the mobility $\mu = \sigma / en$,
\begin{equation}
  \label{eq:effective_medium_assymptot}
  \mu \approx \mu_0 \, \frac{N_c}{2n} \exp\left( \frac{E_f-E^*}{kT} \right) \; .
\end{equation}
The expression~(\ref{eq:effective_medium_assymptot})
is valid when $E_f < E^*$, $kT \ll E^*-E_f$ and $kT \ll \delta$.

According to Eq.~(\ref{eq:effective_medium_assymptot})
one can interpret the transport in the low-temperature
and low-concentration case as thermal activation of electrons
to the energy level $E^*$.
This result is to be compared with that of the percolation theory
given by Eq.~(\ref{eq:mu_percolation1_assymptot}).
Even if we ignore the differences in the pre-exponential factors
we can conclude that the results given by
Eqs.~(\ref{eq:effective_medium_assymptot})
and~(\ref{eq:mu_percolation1_assymptot})
differ by an exponential factor $\propto\exp(-0.52\,\delta/kT)$,
which is essential for strong disorder, $kT \ll \delta$.

However, this result is specific to the present case and
does not imply that the EMA framework
always leads to an exponentially large error
in the case of strong disorder.
As has been shown in several studies,\cite{Nakamura1984,CHEN2006}
one can achieve a better description of the conductivity
within the EMA framework by replacing the spatial dimension $d=3$ with the
inverse of the percolation threshold $1/\vartheta_c\approx 6$.
However, conceptual improvements of the EMA are beyond
the scope of our present study. Instead, in the next Sec.~\ref{sec:results}
we turn to a comparison between the results of percolation theory
from Sec.~\ref{sec:percolation theory} and experimental data.

\section{Comparison with experimental data}
\label{sec:results}

In this section, we show that the percolation approach
developed in Sec.~\ref{sec:percolation theory}
and applied to the random band-edge model
presented in Sec.~\ref{sec:model}
is able to reproduce experimental data on charge transport
in InGaZnO materials.
The main theoretical result to be compared with experimental data is
Eq.~(\ref{eq:sigma_percolation1}) for the conductivity~$\sigma(n,T)$,
as discussed in Sec.~\ref{sec:percolation theory}.
The regional conductivity $\sigma_{\text{region}}(E_m)$
in this equation is given by Eq.~(\ref{eq:sigma_local}),
where the regional density of states $g(E-E_m)$
is taken in the form of Eq.~(\ref{eq:DOS_full})
with $g_c = 10^{21}\, \text{cm}^{-3}\text{eV}^{-3/2}$.
Following Ref.~[\onlinecite{Fishchuk2016}], we take into account the
distribution $G(E_m)$ of the regional positions of the band edge $E_m$
in the form of Eq.~(\ref{eq:DOS_Gauss_E_m}),
and neglect for simplicity the presence of localized states
with energies below $E_m$ by setting $N_m = 0$.
We address experimental data
for the temperature dependencies of conductivity $\sigma(T)$ and
mobility $\mu(T)$ at different concentrations of charge carriers~$n$.
The carrier concentration is changed experimentally
either by varying the doping level,\cite{Kamiya2010_APL}
or by varying the gate voltage
in the field-effect transistors.\cite{Lee2012_APL, Germs2012, Fishchuk2016}

In their pioneering works,\cite{Kamiya2009,Kamiya2010_APL}
Kamiya \emph{et al.}
investigated two series of n-type IGZO films:
crystalline (c-IGZO) and amorphous (a-IGZO).
Samples of c-IGZO are crystalline materials
but they contain inherent disorder due to the statistical distribution of Ga and Zn ions.
Therefore, such materials are to be considered as disordered materials
with respect to charge transport.\cite{Kamiya2010_APL}
In each series of the samples, the conductivity $\sigma(T)$
was measured varying the carrier concentrations~$n$
between
$n < 10^{16}\, \text{cm}^{-3}$ and
$n \sim 10^{20}\,\text{cm}^{-3}$.\cite{Kamiya2010_APL}

\begin{figure}[t]
\includegraphics[width=\linewidth]{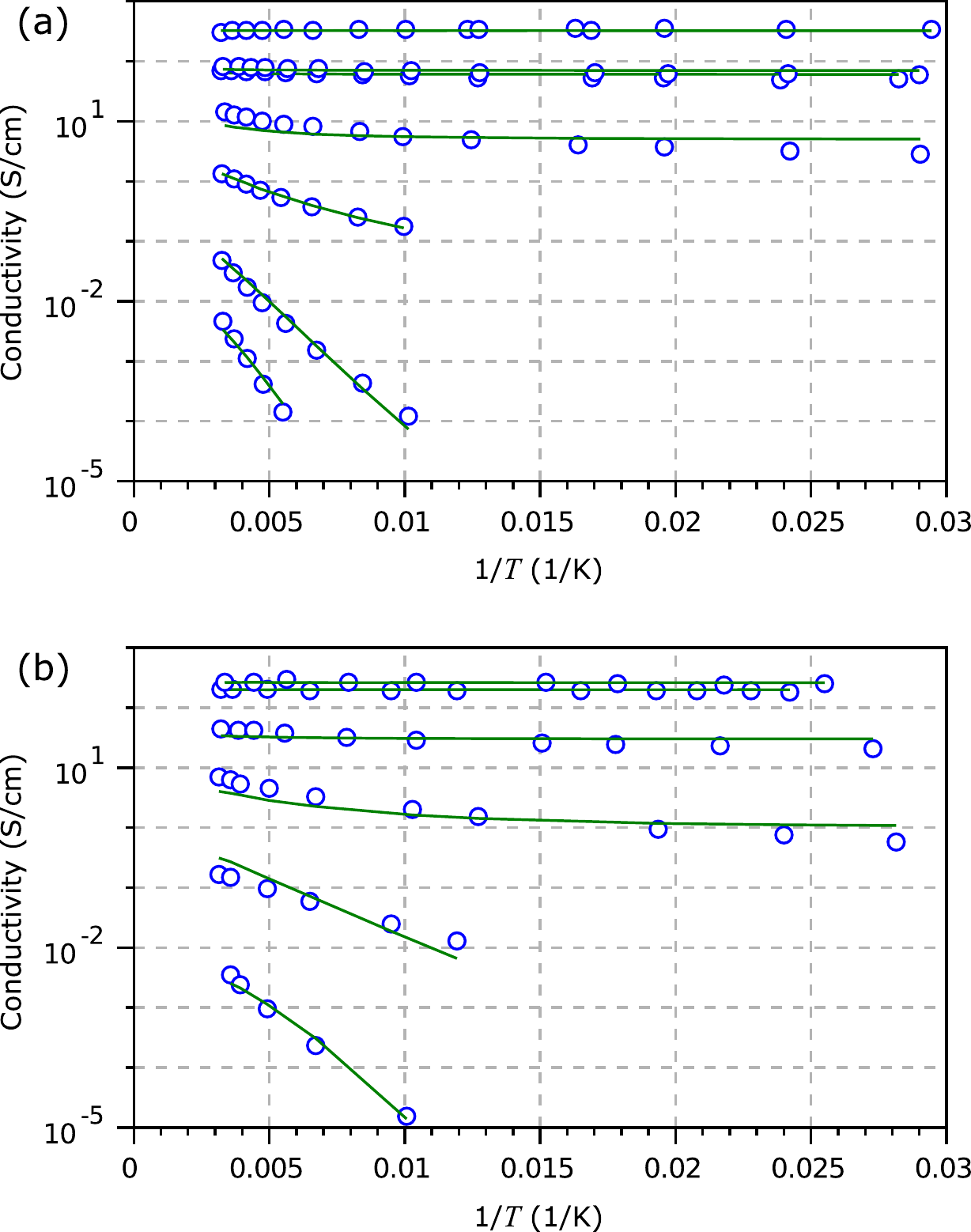}
\caption{Temperature dependence of the conductivity $\sigma(T)$
  in c-IGZO (a) and a-IGZO (b) samples with different carrier concentration~$n$.
  Circles: experimental data.\cite{Kamiya2010_APL}
  Solid lines: fit to Eq.~(\ref{eq:sigma_percolation1}).
  The values of fitting parameters are specified in the text.\label{fig:Kamiya_fitting}}
\end{figure}

Experimental data on the temperature dependencies
of the conductivity $\sigma(T)$ are shown by circles
in Fig.~\ref{fig:Kamiya_fitting}a for c-IGZO
and in Fig.~\ref{fig:Kamiya_fitting}b for a-IGZO.
These data are copied from Fig.~1b,
and Fig.~1d of Ref.~[\onlinecite{Kamiya2010_APL}], respectively.
Theoretical results given by  Eq.~(\ref{eq:sigma_percolation1})
are shown in Fig.~\ref{fig:Kamiya_fitting} by solid lines.
These results are obtained by adjusting the band-edge disorder parameter $\delta$
in Eq.~(\ref{eq:DOS_Gauss_E_m}) and the conduction-band mobility $\mu_0$,
keeping these parameters fixed for \emph{each group of samples}.
Since the values of the carrier concentration~$n$
in different samples were not exactly specified,\cite{Kamiya2010_APL}
we use $n$ as an adjustable parameter. Values for $n$
in the range between $n=10^{15}\, \text{cm}^{-3}$
and $n=6\cdot10^{19}\, \text{cm}^{-3}$
give the best fits, in good agreement
with experimental estimates.\cite{Kamiya2010_APL}
The parameters $\delta$, $\mu_0$, and $n$ are considered
as independent of temperature.
The values of $\delta$ and $\mu_0$ that provide the best fit to the experimental data
are $\delta = 0.057\, \text{eV}$, $\mu_0 = 39\, \text{cm}^2/\text{Vs}$
for c-IGZO, and $\delta = 0.036\, \text{eV}$, $\mu_0 = 47\, \text{cm}^2/\text{Vs}$
for a-IGZO.

\begin{figure}[ht]
\includegraphics[width=\linewidth]{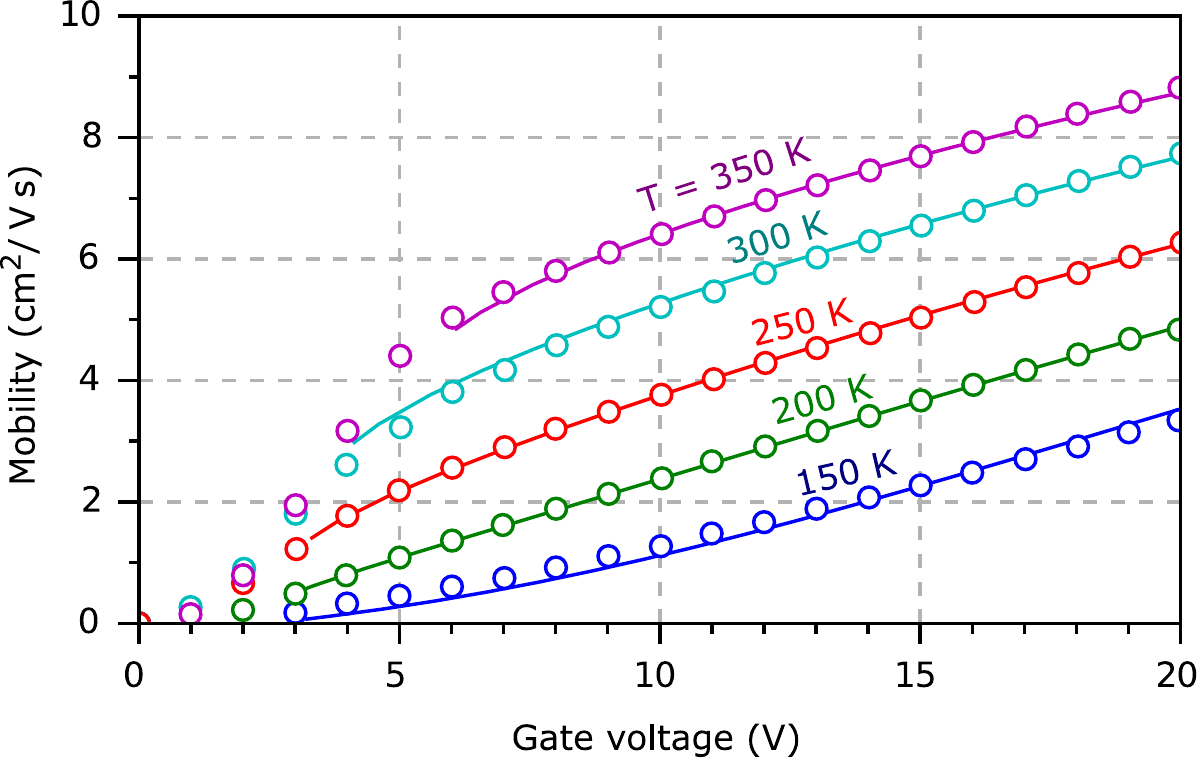}
\caption{Dependence of the carrier mobility $\mu$
  on the gate voltage~$V_g$ at different temperatures.
  Circles: experimental data.\cite{Germs2012}
  Solid lines: fit $\mu(n,T) = \sigma(n,T)/(en$), where $\sigma(n,T)$
  is given by Eq.~(\ref{eq:sigma_percolation1}).
  The values of fitting parameters are specified in the text.\label{fig:Germs_fitting}}
\end{figure}

Another set of experimental data to be compared with theoretical predictions
is related to the carrier mobility $\mu(n,T)$ measured in thin-film transistors
with IGZO channels.\cite{Germs2012, Fishchuk2016}
In Fig.~\ref{fig:Germs_fitting}, experimental data for the dependencies of $\mu$
on the gate voltage $V_g$ at different temperatures
are reproduced from Ref.~[\onlinecite{Germs2012}] as depicted by circles.
Solid lines are the theoretical results for $\mu(n,T) = \sigma(n,T)/(en)$,
where $\sigma(n,T)$ is obtained from Eq.~(\ref{eq:sigma_percolation1}).
The carrier concentration~$n$ is assumed linearly dependent on the gate voltage $V_g$,
\begin{equation}
\label{eq:n_versus_gate_voltage}
n = \lambda [V_g - V^*(T)] \; ,
\end{equation}
where the proportionality constant~$\lambda$ serves as a fitting parameter.
It depends on the relative capacitance between the gate and the channel
and on the thickness of the electron accumulation layer.
The value $\lambda = 9.11 \cdot 10^{16}\, \text{cm}^{-3}\text{V}^{-1}$
gives the best fit. Following Ref.~[\onlinecite{Germs2012}],
the experimental gate voltage is counted from its threshold value.
The flat-band voltage $V^*$ is also considered as an adjustable parameter.
The values $V^* = 2.69\, \text{V}, 2.32 \, \text{V}, 2.79\, \text{V},
3.61\, \text{V}, 5.57\, \text{V}$ are used for
$T = 150\, \text{K}, 200\, \text{K}, 250\, \text{K}, 300\, \text{K}, 350\, \text{K}$,
respectively.
The best fits to the experimental data in Fig.~\ref{fig:Germs_fitting} are
achieved by choosing the band-edge disorder parameter
$\delta = 0.063\, \text{eV}$ in Eq.~(\ref{eq:DOS_Gauss_E_m})
and the conduction-band mobility $\mu_0 = 30\, \text{cm}^2/\text{Vs}$.

\begin{figure}[ht]
\includegraphics[width=\linewidth]{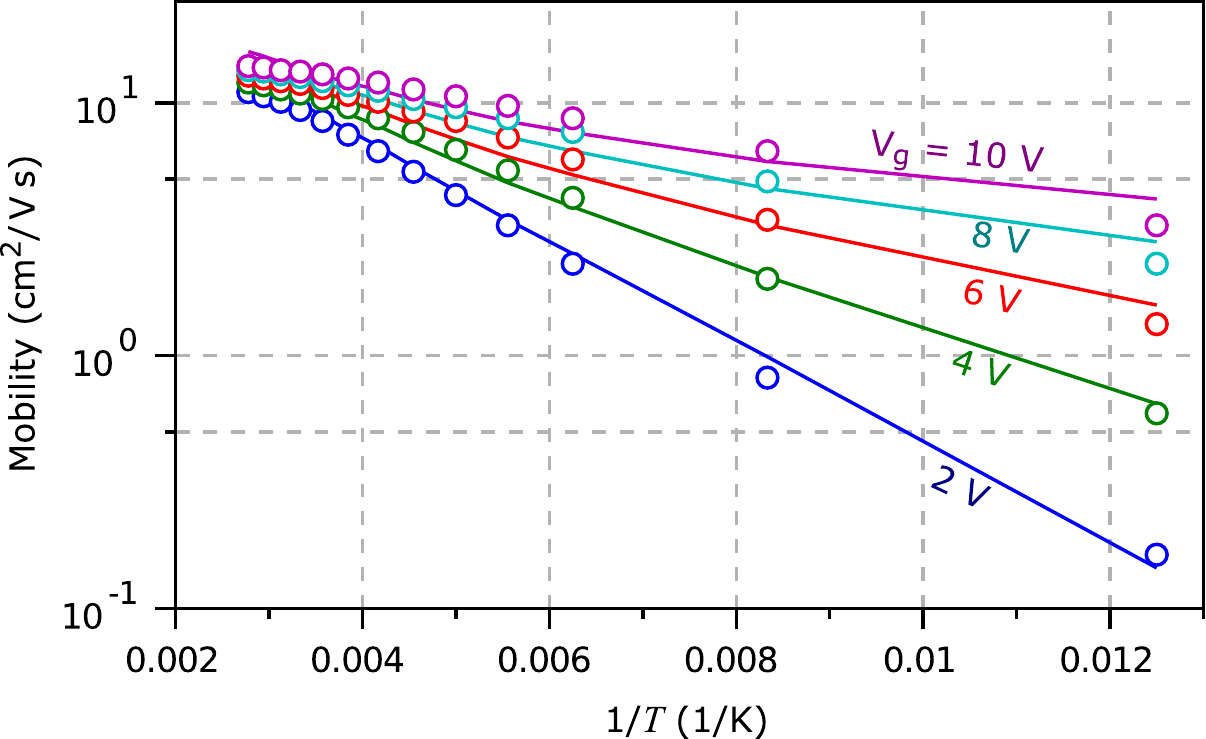}
\caption{Dependence of the carrier mobility $\mu(V_g,T)$
  on temperature at different gate voltages $V_g$. Circles:
  experimental data.\cite{Fishchuk2016}
  Solid lines: fit of $\mu(n,T) = \sigma(n,T)/(en)$, where $\sigma(n,T)$
  is given by Eq.~(\ref{eq:sigma_percolation1}).
  The values of fitting parameters are specified in the text.\label{fig:Fishchuk_fitting}}
\end{figure}

Experimental data on the carrier mobility in a-IGZO thin-film transistors,
analogous to those in Fig.~\ref{fig:Germs_fitting},
were obtained by Fishchuk \textit{et al.}~\cite{Fishchuk2016}
who converted the data into $\mu(T)$
at different gate voltages $V_g$.
The data are shown by circles in Fig.~\ref{fig:Fishchuk_fitting}.
Solid lines are fits to Eq.~(\ref{eq:sigma_percolation1}).
The temperature dependence of $V^*$ in Eq.~(\ref{eq:n_versus_gate_voltage})
is taken from Ref.~[\onlinecite{Fishchuk2016}], $V^*(T) = -1.61\, \text{V}
+ 109\, \text{V}/(T/\text{K})$.
The value $\lambda = 1.32\cdot 10^{17}\, \text{cm}^{-3}\text{V}^{-1}$
gives the best fit.
The best agreement with experimental data in Fig.~\ref{fig:Fishchuk_fitting} is
achieved by choosing the band-edge disorder parameter
$\delta = 0.05\, \text{eV}$ in Eq.~(\ref{eq:DOS_Gauss_E_m})
and the conduction-band mobility $\mu_0 = 36\, \text{cm}^2/\text{(Vs)}$.
These values are close to the values $\delta = 0.04\, \text{eV}$
  and $\mu_0 = 22\, \text{cm}^2$/\text{(Vs)}
  obtained by Fishchuk \textit{et al.}~\cite{Fishchuk2016}
  from a comparison of their experimental data and
  their theory based on the EMA.
  This evidences that there is not much difference
  between the results of the percolation theory and
  those of the EMA for the range of parameters $\delta$, $kT$, $n$, and $\mu_0$
  relevant to the experimental situation studied in Ref.~[\onlinecite{Fishchuk2016}].

For a fairly small range of parameters,
$36\, \text{meV}<\delta<63\, \text{meV}$
and $30\, \text{cm}^2/\text{(Vs)}<\mu_0<47\, \text{cm}^2/\text{(Vs)}$
for the band-edge disorder parameter
$\delta$ in Eq.~(\ref{eq:DOS_Gauss_E_m}) and the
conduction-band mobility $\mu_0$,
percolation theory reliably reproduces
different sets of experimental data in IGZO materials
over a broad range of temperatures and charge carrier densities.
%Moreover, these parameter values look fairly reasonable
%for charge transport in disordered oxide semiconductors.

\section{Conclusions}
\label{sec:conclusions}

Theoretical approach
based on the percolation theory is developed
to describe charge transport in amorphous oxide semiconductors
in the framework of the random band-edge model
that takes into account the effect of disorder
on the regional position of the band edge $E_m$.
For the case of a Gaussian distribution for $E_m$,
the superiority of the percolation approach is proven in comparison
with previously used averaging schemes.
Our percolation approach reproduces experimental data on charge transport
in IGZO materials obtained by several groups. The comparison between theoretical results and experimental data reveals the energy scale of disorder in such materials.

%The comparison between theory and experiments
%reveals that the width of the band-edge disorder~$\delta$
%  and the bare electron mobility are the crucial parameters
%  for charge transport in these materials.

\begin{acknowledgments}

  The authors are indebted to Prof.\ A.\ Kadashchuk and to Prof.\ H.\ Hosono
  for bringing their attention to the problem of charge transport in AOS.
  Financial support of the Deutsche Forschungsgemeinschaft (GRK 1782)
is gratefully acknowledged.
\end{acknowledgments}

\bibliography{IGZO}

\begin{thebibliography}{34}%
\makeatletter
\providecommand \@ifxundefined [1]{%
 \@ifx{#1\undefined}
}%
\providecommand \@ifnum [1]{%
 \ifnum #1\expandafter \@firstoftwo
 \else \expandafter \@secondoftwo
 \fi
}%
\providecommand \@ifx [1]{%
 \ifx #1\expandafter \@firstoftwo
 \else \expandafter \@secondoftwo
 \fi
}%
\providecommand \natexlab [1]{#1}%
\providecommand \enquote  [1]{``#1''}%
\providecommand \bibnamefont  [1]{#1}%
\providecommand \bibfnamefont [1]{#1}%
\providecommand \citenamefont [1]{#1}%
\providecommand \href@noop [0]{\@secondoftwo}%
\providecommand \href [0]{\begingroup \@sanitize@url \@href}%
\providecommand \@href[1]{\@@startlink{#1}\@@href}%
\providecommand \@@href[1]{\endgroup#1\@@endlink}%
\providecommand \@sanitize@url [0]{\catcode `\\12\catcode `\$12\catcode
  `\&12\catcode `\#12\catcode `\^12\catcode `\_12\catcode `\%12\relax}%
\providecommand \@@startlink[1]{}%
\providecommand \@@endlink[0]{}%
\providecommand \url  [0]{\begingroup\@sanitize@url \@url }%
\providecommand \@url [1]{\endgroup\@href {#1}{\urlprefix }}%
\providecommand \urlprefix  [0]{URL }%
\providecommand \Eprint [0]{\href }%
\providecommand \doibase [0]{http://dx.doi.org/}%
\providecommand \selectlanguage [0]{\@gobble}%
\providecommand \bibinfo  [0]{\@secondoftwo}%
\providecommand \bibfield  [0]{\@secondoftwo}%
\providecommand \translation [1]{[#1]}%
\providecommand \BibitemOpen [0]{}%
\providecommand \bibitemStop [0]{}%
\providecommand \bibitemNoStop [0]{.\EOS\space}%
\providecommand \EOS [0]{\spacefactor3000\relax}%
\providecommand \BibitemShut  [1]{\csname bibitem#1\endcsname}%
\let\auto@bib@innerbib\@empty
%</preamble>
\bibitem [{\citenamefont {Nomura}\ \emph {et~al.}(2004)\citenamefont {Nomura},
  \citenamefont {Ohta}, \citenamefont {Takagi}, \citenamefont {Kamiya},
  \citenamefont {Hirano},\ and\ \citenamefont {Hosono}}]{Nomura2004_Nature}%
  \BibitemOpen
  \bibfield  {author} {\bibinfo {author} {\bibfnamefont {K.}~\bibnamefont
  {Nomura}}, \bibinfo {author} {\bibfnamefont {H.}~\bibnamefont {Ohta}},
  \bibinfo {author} {\bibfnamefont {A.}~\bibnamefont {Takagi}}, \bibinfo
  {author} {\bibfnamefont {T.}~\bibnamefont {Kamiya}}, \bibinfo {author}
  {\bibfnamefont {M.}~\bibnamefont {Hirano}}, \ and\ \bibinfo {author}
  {\bibfnamefont {H.}~\bibnamefont {Hosono}},\ }\href@noop {} {\bibfield
  {journal} {\bibinfo  {journal} {Nature}\ }\textbf {\bibinfo {volume} {432}},\
  \bibinfo {pages} {488} (\bibinfo {year} {2004})}\BibitemShut {NoStop}%
\bibitem [{\citenamefont {Kamiya}\ \emph {et~al.}(2009)\citenamefont {Kamiya},
  \citenamefont {Nomura},\ and\ \citenamefont {Hosono}}]{Kamiya2009}%
  \BibitemOpen
  \bibfield  {author} {\bibinfo {author} {\bibfnamefont {T.}~\bibnamefont
  {Kamiya}}, \bibinfo {author} {\bibfnamefont {K.}~\bibnamefont {Nomura}}, \
  and\ \bibinfo {author} {\bibfnamefont {H.}~\bibnamefont {Hosono}},\
  }\href@noop {} {\bibfield  {journal} {\bibinfo  {journal} {J. Disp.
  Technol.}\ }\textbf {\bibinfo {volume} {5}},\ \bibinfo {pages} {462}
  (\bibinfo {year} {2009})}\BibitemShut {NoStop}%
\bibitem [{\citenamefont {Takagi}\ \emph {et~al.}(2005)\citenamefont {Takagi},
  \citenamefont {Nomura}, \citenamefont {Ohta}, \citenamefont {Yanagi},
  \citenamefont {Kamiya}, \citenamefont {Hirano},\ and\ \citenamefont
  {Hosono}}]{TAKAGI2005}%
  \BibitemOpen
  \bibfield  {author} {\bibinfo {author} {\bibfnamefont {A.}~\bibnamefont
  {Takagi}}, \bibinfo {author} {\bibfnamefont {K.}~\bibnamefont {Nomura}},
  \bibinfo {author} {\bibfnamefont {H.}~\bibnamefont {Ohta}}, \bibinfo {author}
  {\bibfnamefont {H.}~\bibnamefont {Yanagi}}, \bibinfo {author} {\bibfnamefont
  {T.}~\bibnamefont {Kamiya}}, \bibinfo {author} {\bibfnamefont
  {M.}~\bibnamefont {Hirano}}, \ and\ \bibinfo {author} {\bibfnamefont
  {H.}~\bibnamefont {Hosono}},\ }\href {\doibase
  https://doi.org/10.1016/j.tsf.2004.11.223} {\bibfield  {journal} {\bibinfo
  {journal} {Thin Solid Films}\ }\textbf {\bibinfo {volume} {486}},\ \bibinfo
  {pages} {38} (\bibinfo {year} {2005})}\BibitemShut {NoStop}%
\bibitem [{\citenamefont {Kamiya}\ \emph {et~al.}(2010)\citenamefont {Kamiya},
  \citenamefont {Nomura},\ and\ \citenamefont {Hosono}}]{Kamiya2010_APL}%
  \BibitemOpen
  \bibfield  {author} {\bibinfo {author} {\bibfnamefont {T.}~\bibnamefont
  {Kamiya}}, \bibinfo {author} {\bibfnamefont {K.}~\bibnamefont {Nomura}}, \
  and\ \bibinfo {author} {\bibfnamefont {H.}~\bibnamefont {Hosono}},\ }\href
  {https://doi.org/10.1063/1.3364131} {\bibfield  {journal} {\bibinfo
  {journal} {Appl. Phys. Lett.}\ }\textbf {\bibinfo {volume} {96}},\ \bibinfo
  {pages} {122103} (\bibinfo {year} {2010})}\BibitemShut {NoStop}%
\bibitem [{\citenamefont {Kimura}\ \emph {et~al.}(2010)\citenamefont {Kimura},
  \citenamefont {Kamiya}, \citenamefont {Nakanishi}, \citenamefont {Nomura},\
  and\ \citenamefont {Hosono}}]{Kimura2010_APL}%
  \BibitemOpen
  \bibfield  {author} {\bibinfo {author} {\bibfnamefont {M.}~\bibnamefont
  {Kimura}}, \bibinfo {author} {\bibfnamefont {T.}~\bibnamefont {Kamiya}},
  \bibinfo {author} {\bibfnamefont {T.}~\bibnamefont {Nakanishi}}, \bibinfo
  {author} {\bibfnamefont {K.}~\bibnamefont {Nomura}}, \ and\ \bibinfo {author}
  {\bibfnamefont {H.}~\bibnamefont {Hosono}},\ }\href {\doibase
  10.1063/1.3455072} {\bibfield  {journal} {\bibinfo  {journal} {Appl. Phys.
  Lett.}\ }\textbf {\bibinfo {volume} {96}},\ \bibinfo {pages} {262105}
  (\bibinfo {year} {2010})}\BibitemShut {NoStop}%
\bibitem [{\citenamefont {Sze}(1981)}]{Sze1981}%
  \BibitemOpen
  \bibfield  {author} {\bibinfo {author} {\bibfnamefont {S.~M.}\ \bibnamefont
  {Sze}},\ }\href@noop {} {\emph {\bibinfo {title} {Physics of Semiconductor
  Devices}}},\ \bibinfo {edition} {2nd}\ ed.\ (\bibinfo  {publisher} {Wiley,
  New York},\ \bibinfo {year} {1981})\BibitemShut {NoStop}%
\bibitem [{\citenamefont {Adler}\ \emph {et~al.}(1973)\citenamefont {Adler},
  \citenamefont {Flora},\ and\ \citenamefont {Sentuna}}]{Adler1973}%
  \BibitemOpen
  \bibfield  {author} {\bibinfo {author} {\bibfnamefont {D.}~\bibnamefont
  {Adler}}, \bibinfo {author} {\bibfnamefont {L.~P.}\ \bibnamefont {Flora}}, \
  and\ \bibinfo {author} {\bibfnamefont {S.~D.}\ \bibnamefont {Sentuna}},\
  }\href@noop {} {\bibfield  {journal} {\bibinfo  {journal} {Solid State
  Commun.}\ }\textbf {\bibinfo {volume} {12}},\ \bibinfo {pages} {9} (\bibinfo
  {year} {1973})}\BibitemShut {NoStop}%
\bibitem [{\citenamefont {Park}\ \emph {et~al.}(2010)\citenamefont {Park},
  \citenamefont {Jeon}, \citenamefont {Lee}, \citenamefont {Kim}, \citenamefont
  {Kim}, \citenamefont {Song}, \citenamefont {Park}, \citenamefont {Park},
  \citenamefont {Kim}, \citenamefont {Kim},\ and\ \citenamefont
  {Kim}}]{Park2010}%
  \BibitemOpen
  \bibfield  {author} {\bibinfo {author} {\bibfnamefont {J.-H.}\ \bibnamefont
  {Park}}, \bibinfo {author} {\bibfnamefont {K.}~\bibnamefont {Jeon}}, \bibinfo
  {author} {\bibfnamefont {S.}~\bibnamefont {Lee}}, \bibinfo {author}
  {\bibfnamefont {S.}~\bibnamefont {Kim}}, \bibinfo {author} {\bibfnamefont
  {S.}~\bibnamefont {Kim}}, \bibinfo {author} {\bibfnamefont {I.}~\bibnamefont
  {Song}}, \bibinfo {author} {\bibfnamefont {J.}~\bibnamefont {Park}}, \bibinfo
  {author} {\bibfnamefont {Y.}~\bibnamefont {Park}}, \bibinfo {author}
  {\bibfnamefont {C.~J.}\ \bibnamefont {Kim}}, \bibinfo {author} {\bibfnamefont
  {D.~M.}\ \bibnamefont {Kim}}, \ and\ \bibinfo {author} {\bibfnamefont
  {D.~H.}\ \bibnamefont {Kim}},\ }\href@noop {} {\bibfield  {journal} {\bibinfo
   {journal} {J. Electrochem. Soc.}\ }\textbf {\bibinfo {volume} {157}},\
  \bibinfo {pages} {H272} (\bibinfo {year} {2010})}\BibitemShut {NoStop}%
\bibitem [{\citenamefont {Lee}\ \emph {et~al.}(2010)\citenamefont {Lee},
  \citenamefont {Park}, \citenamefont {Kim}, \citenamefont {Jeon},
  \citenamefont {Jeon}, \citenamefont {Park}, \citenamefont {Park},
  \citenamefont {Song}, \citenamefont {Kim}, \citenamefont {Y.Park},
  \citenamefont {Kim},\ and\ \citenamefont {Kim}}]{LeeParkKim2010}%
  \BibitemOpen
  \bibfield  {author} {\bibinfo {author} {\bibfnamefont {S.}~\bibnamefont
  {Lee}}, \bibinfo {author} {\bibfnamefont {S.}~\bibnamefont {Park}}, \bibinfo
  {author} {\bibfnamefont {S.}~\bibnamefont {Kim}}, \bibinfo {author}
  {\bibfnamefont {Y.}~\bibnamefont {Jeon}}, \bibinfo {author} {\bibfnamefont
  {K.}~\bibnamefont {Jeon}}, \bibinfo {author} {\bibfnamefont {J.-H.}\
  \bibnamefont {Park}}, \bibinfo {author} {\bibfnamefont {J.}~\bibnamefont
  {Park}}, \bibinfo {author} {\bibfnamefont {I.}~\bibnamefont {Song}}, \bibinfo
  {author} {\bibfnamefont {C.~J.}\ \bibnamefont {Kim}}, \bibinfo {author}
  {\bibnamefont {Y.Park}}, \bibinfo {author} {\bibfnamefont {D.~M.}\
  \bibnamefont {Kim}}, \ and\ \bibinfo {author} {\bibfnamefont {D.~H.}\
  \bibnamefont {Kim}},\ }\href@noop {} {\bibfield  {journal} {\bibinfo
  {journal} {IEEE Electron Device Lett.}\ }\textbf {\bibinfo {volume} {31}},\
  \bibinfo {pages} {231} (\bibinfo {year} {2010})}\BibitemShut {NoStop}%
\bibitem [{\citenamefont {Mott}\ and\ \citenamefont {Davis}(1979)}]{Mott1979}%
  \BibitemOpen
  \bibfield  {author} {\bibinfo {author} {\bibfnamefont {N.~F.}\ \bibnamefont
  {Mott}}\ and\ \bibinfo {author} {\bibfnamefont {E.~A.}\ \bibnamefont
  {Davis}},\ }\href@noop {} {\emph {\bibinfo {title} {Electronic Processes in
  Non-Crystalline Materials}}},\ \bibinfo {edition} {2nd}\ ed.\ (\bibinfo
  {publisher} {Clarendon Press, Oxford},\ \bibinfo {year} {1979})\BibitemShut
  {NoStop}%
\bibitem [{\citenamefont {Overhof}\ and\ \citenamefont
  {Thomas}(1989)}]{Thomas1989}%
  \BibitemOpen
  \bibfield  {author} {\bibinfo {author} {\bibfnamefont {H.}~\bibnamefont
  {Overhof}}\ and\ \bibinfo {author} {\bibfnamefont {P.}~\bibnamefont
  {Thomas}},\ }\href@noop {} {\emph {\bibinfo {title} {Electronic Transport in
  Hydrogenated Amorphous Semiconductors}}}\ (\bibinfo  {publisher} {Springer,
  Heidelberg},\ \bibinfo {year} {1989})\BibitemShut {NoStop}%
\bibitem [{\citenamefont {Street}(1991)}]{Street1991}%
  \BibitemOpen
  \bibfield  {author} {\bibinfo {author} {\bibfnamefont {R.~A.}\ \bibnamefont
  {Street}},\ }\href@noop {} {\emph {\bibinfo {title} {Hydrogenated Amorphous
  Silicon}}},\ Cambridge Solid State Science Series\ (\bibinfo  {publisher}
  {Cambridge University Press},\ \bibinfo {year} {1991})\BibitemShut {NoStop}%
\bibitem [{\citenamefont {Baranovski}(2006)}]{Baranovski2006}%
  \BibitemOpen
  \bibinfo {editor} {\bibfnamefont {S.}~\bibnamefont {Baranovski}},\ ed.,\
  \href@noop {} {\emph {\bibinfo {title} {Charge Transport in Disordered Solids
  with Applications in Electronics}}}\ (\bibinfo  {publisher} {John Wiley \&
  Sons, Ltd, Chichester},\ \bibinfo {year} {2006})\BibitemShut {NoStop}%
\bibitem [{\citenamefont {Semeniuk}\ \emph {et~al.}(2017)\citenamefont
  {Semeniuk}, \citenamefont {Juska}, \citenamefont {Oelerich}, \citenamefont
  {Jandieri}, \citenamefont {Baranovskii},\ and\ \citenamefont
  {Reznik}}]{Smeniuk2017}%
  \BibitemOpen
  \bibfield  {author} {\bibinfo {author} {\bibfnamefont {O.}~\bibnamefont
  {Semeniuk}}, \bibinfo {author} {\bibfnamefont {G.}~\bibnamefont {Juska}},
  \bibinfo {author} {\bibfnamefont {J.~O.}\ \bibnamefont {Oelerich}}, \bibinfo
  {author} {\bibfnamefont {K.}~\bibnamefont {Jandieri}}, \bibinfo {author}
  {\bibfnamefont {S.~D.}\ \bibnamefont {Baranovskii}}, \ and\ \bibinfo {author}
  {\bibfnamefont {A.}~\bibnamefont {Reznik}},\ }\href@noop {} {\bibfield
  {journal} {\bibinfo  {journal} {J. Phys. D}\ }\textbf {\bibinfo {volume}
  {50}},\ \bibinfo {pages} {035103} (\bibinfo {year} {2017})}\BibitemShut
  {NoStop}%
\bibitem [{\citenamefont {Lee}\ \emph {et~al.}(2011)\citenamefont {Lee},
  \citenamefont {Ghaffarzadeh}, \citenamefont {Nathan}, \citenamefont
  {Robertson}, \citenamefont {Jeon}, \citenamefont {Kim}, \citenamefont
  {Song},\ and\ \citenamefont {Chung}}]{Lee2011_APL}%
  \BibitemOpen
  \bibfield  {author} {\bibinfo {author} {\bibfnamefont {S.}~\bibnamefont
  {Lee}}, \bibinfo {author} {\bibfnamefont {K.}~\bibnamefont {Ghaffarzadeh}},
  \bibinfo {author} {\bibfnamefont {A.}~\bibnamefont {Nathan}}, \bibinfo
  {author} {\bibfnamefont {J.}~\bibnamefont {Robertson}}, \bibinfo {author}
  {\bibfnamefont {S.}~\bibnamefont {Jeon}}, \bibinfo {author} {\bibfnamefont
  {C.}~\bibnamefont {Kim}}, \bibinfo {author} {\bibfnamefont {I.-H.}\
  \bibnamefont {Song}}, \ and\ \bibinfo {author} {\bibfnamefont {U.-I.}\
  \bibnamefont {Chung}},\ }\href {\doibase 10.1063/1.3589371} {\bibfield
  {journal} {\bibinfo  {journal} {Appl. Phys. Lett.}\ }\textbf {\bibinfo
  {volume} {98}},\ \bibinfo {pages} {203508} (\bibinfo {year}
  {2011})}\BibitemShut {NoStop}%
\bibitem [{\citenamefont {Lee}\ and\ \citenamefont
  {Nathan}(2012)}]{Lee2012_APL}%
  \BibitemOpen
  \bibfield  {author} {\bibinfo {author} {\bibfnamefont {S.}~\bibnamefont
  {Lee}}\ and\ \bibinfo {author} {\bibfnamefont {A.}~\bibnamefont {Nathan}},\
  }\href {\doibase 10.1063/1.4751861} {\bibfield  {journal} {\bibinfo
  {journal} {Appl. Phys. Lett.}\ }\textbf {\bibinfo {volume} {101}},\ \bibinfo
  {pages} {113502} (\bibinfo {year} {2012})}\BibitemShut {NoStop}%
\bibitem [{\citenamefont {Shklovskii}\ and\ \citenamefont
  {Efros}(1984)}]{Shklovskii1984}%
  \BibitemOpen
  \bibfield  {author} {\bibinfo {author} {\bibfnamefont {B.~I.}\ \bibnamefont
  {Shklovskii}}\ and\ \bibinfo {author} {\bibfnamefont {A.~L.}\ \bibnamefont
  {Efros}},\ }\href@noop {} {\emph {\bibinfo {title} {Electronic Properties of
  Doped Semiconductors}}}\ (\bibinfo  {publisher} {Springer, Berlin},\ \bibinfo
  {year} {1984})\BibitemShut {NoStop}%
\bibitem [{\citenamefont {Germs}\ \emph {et~al.}(2012)\citenamefont {Germs},
  \citenamefont {Adriaans}, \citenamefont {Tripathi}, \citenamefont {Roelofs},
  \citenamefont {Cobb}, \citenamefont {Janssen}, \citenamefont {Gelinck},\ and\
  \citenamefont {Kemerink}}]{Germs2012}%
  \BibitemOpen
  \bibfield  {author} {\bibinfo {author} {\bibfnamefont {W.~C.}\ \bibnamefont
  {Germs}}, \bibinfo {author} {\bibfnamefont {W.~H.}\ \bibnamefont {Adriaans}},
  \bibinfo {author} {\bibfnamefont {A.~K.}\ \bibnamefont {Tripathi}}, \bibinfo
  {author} {\bibfnamefont {W.~S.~C.}\ \bibnamefont {Roelofs}}, \bibinfo
  {author} {\bibfnamefont {B.}~\bibnamefont {Cobb}}, \bibinfo {author}
  {\bibfnamefont {R.~A.~J.}\ \bibnamefont {Janssen}}, \bibinfo {author}
  {\bibfnamefont {G.~H.}\ \bibnamefont {Gelinck}}, \ and\ \bibinfo {author}
  {\bibfnamefont {M.}~\bibnamefont {Kemerink}},\ }\href {\doibase
  10.1103/PhysRevB.86.155319} {\bibfield  {journal} {\bibinfo  {journal} {Phys.
  Rev. B}\ }\textbf {\bibinfo {volume} {86}},\ \bibinfo {pages} {155319}
  (\bibinfo {year} {2012})}\BibitemShut {NoStop}%
\bibitem [{\citenamefont {Gr\"unewald}\ and\ \citenamefont
  {Thomas}(1979)}]{Gruenewald1979}%
  \BibitemOpen
  \bibfield  {author} {\bibinfo {author} {\bibfnamefont {M.}~\bibnamefont
  {Gr\"unewald}}\ and\ \bibinfo {author} {\bibfnamefont {P.}~\bibnamefont
  {Thomas}},\ }\href@noop {} {\bibfield  {journal} {\bibinfo  {journal} {phys.
  stat. sol. (b)}\ }\textbf {\bibinfo {volume} {94}},\ \bibinfo {pages} {125}
  (\bibinfo {year} {1979})}\BibitemShut {NoStop}%
\bibitem [{\citenamefont {Monroe}(1985)}]{Monroe1985}%
  \BibitemOpen
  \bibfield  {author} {\bibinfo {author} {\bibfnamefont {D.}~\bibnamefont
  {Monroe}},\ }\href@noop {} {\bibfield  {journal} {\bibinfo  {journal} {Phys.
  Rev. Lett.}\ }\textbf {\bibinfo {volume} {54}},\ \bibinfo {pages} {146}
  (\bibinfo {year} {1985})}\BibitemShut {NoStop}%
\bibitem [{\citenamefont {Baranovskii}\ \emph {et~al.}(1995)\citenamefont
  {Baranovskii}, \citenamefont {Thomas},\ and\ \citenamefont
  {Adriaenssens}}]{Baranovskii1995}%
  \BibitemOpen
  \bibfield  {author} {\bibinfo {author} {\bibfnamefont {S.~D.}\ \bibnamefont
  {Baranovskii}}, \bibinfo {author} {\bibfnamefont {P.}~\bibnamefont {Thomas}},
  \ and\ \bibinfo {author} {\bibfnamefont {G.~J.}\ \bibnamefont
  {Adriaenssens}},\ }\href@noop {} {\bibfield  {journal} {\bibinfo  {journal}
  {J. Non-Cryst. Solids}\ }\textbf {\bibinfo {volume} {190}},\ \bibinfo {pages}
  {283} (\bibinfo {year} {1995})}\BibitemShut {NoStop}%
\bibitem [{\citenamefont {Baranovskii}\ \emph {et~al.}(1997)\citenamefont
  {Baranovskii}, \citenamefont {Faber}, \citenamefont {Hensel},\ and\
  \citenamefont {Thomas}}]{Baranovskii1997}%
  \BibitemOpen
  \bibfield  {author} {\bibinfo {author} {\bibfnamefont {S.~D.}\ \bibnamefont
  {Baranovskii}}, \bibinfo {author} {\bibfnamefont {T.}~\bibnamefont {Faber}},
  \bibinfo {author} {\bibfnamefont {F.}~\bibnamefont {Hensel}}, \ and\ \bibinfo
  {author} {\bibfnamefont {P.}~\bibnamefont {Thomas}},\ }\href@noop {}
  {\bibfield  {journal} {\bibinfo  {journal} {J. Phys. C}\ }\textbf {\bibinfo
  {volume} {9}},\ \bibinfo {pages} {2699} (\bibinfo {year} {1997})}\BibitemShut
  {NoStop}%
\bibitem [{\citenamefont {Baranovskii}\ \emph {et~al.}(2000)\citenamefont
  {Baranovskii}, \citenamefont {Cordes}, \citenamefont {Hensel},\ and\
  \citenamefont {Leising}}]{Baranovskii2000}%
  \BibitemOpen
  \bibfield  {author} {\bibinfo {author} {\bibfnamefont {S.~D.}\ \bibnamefont
  {Baranovskii}}, \bibinfo {author} {\bibfnamefont {H.}~\bibnamefont {Cordes}},
  \bibinfo {author} {\bibfnamefont {F.}~\bibnamefont {Hensel}}, \ and\ \bibinfo
  {author} {\bibfnamefont {G.}~\bibnamefont {Leising}},\ }\href@noop {}
  {\bibfield  {journal} {\bibinfo  {journal} {Phys. Rev. B}\ }\textbf {\bibinfo
  {volume} {62}},\ \bibinfo {pages} {7934} (\bibinfo {year}
  {2000})}\BibitemShut {NoStop}%
\bibitem [{\citenamefont {Baranovskii}\ \emph {et~al.}(2002)\citenamefont
  {Baranovskii}, \citenamefont {Zvyagin}, \citenamefont {Cordes}, \citenamefont
  {Yamasaki},\ and\ \citenamefont {Thomas}}]{Baranovskii2002a}%
  \BibitemOpen
  \bibfield  {author} {\bibinfo {author} {\bibfnamefont {S.~D.}\ \bibnamefont
  {Baranovskii}}, \bibinfo {author} {\bibfnamefont {I.~P.}\ \bibnamefont
  {Zvyagin}}, \bibinfo {author} {\bibfnamefont {H.}~\bibnamefont {Cordes}},
  \bibinfo {author} {\bibfnamefont {S.}~\bibnamefont {Yamasaki}}, \ and\
  \bibinfo {author} {\bibfnamefont {P.}~\bibnamefont {Thomas}},\ }\href@noop {}
  {\bibfield  {journal} {\bibinfo  {journal} {phys. stat. sol. (b)}\ }\textbf
  {\bibinfo {volume} {230}},\ \bibinfo {pages} {281} (\bibinfo {year}
  {2002})}\BibitemShut {NoStop}%
\bibitem [{\citenamefont {Baranovskii}(2014)}]{Baranovskii2014}%
  \BibitemOpen
  \bibfield  {author} {\bibinfo {author} {\bibfnamefont {S.~D.}\ \bibnamefont
  {Baranovskii}},\ }\href@noop {} {\bibfield  {journal} {\bibinfo  {journal}
  {phys. stat. sol. (b)}\ }\textbf {\bibinfo {volume} {251}},\ \bibinfo {pages}
  {487} (\bibinfo {year} {2014})}\BibitemShut {NoStop}%
\bibitem [{\citenamefont {Nenashev}\ \emph {et~al.}(2015)\citenamefont
  {Nenashev}, \citenamefont {Oelerich},\ and\ \citenamefont
  {Baranovskii}}]{Nenashev_Topical_2015}%
  \BibitemOpen
  \bibfield  {author} {\bibinfo {author} {\bibfnamefont {A.~V.}\ \bibnamefont
  {Nenashev}}, \bibinfo {author} {\bibfnamefont {J.~O.}\ \bibnamefont
  {Oelerich}}, \ and\ \bibinfo {author} {\bibfnamefont {S.~D.}\ \bibnamefont
  {Baranovskii}},\ }\href@noop {} {\bibfield  {journal} {\bibinfo  {journal}
  {J. Phys.: Condens. Matter}\ }\textbf {\bibinfo {volume} {27}},\ \bibinfo
  {pages} {093201} (\bibinfo {year} {2015})}\BibitemShut {NoStop}%
\bibitem [{\citenamefont {Fishchuk}\ \emph {et~al.}(2016)\citenamefont
  {Fishchuk}, \citenamefont {Kadashchuk}, \citenamefont {Bhoolokam},
  \citenamefont {de~Jamblinne~de Meux}, \citenamefont {Pourtois}, \citenamefont
  {Gavrilyuk}, \citenamefont {K\"ohler}, \citenamefont {B\"assler},
  \citenamefont {Heremans},\ and\ \citenamefont {Genoe}}]{Fishchuk2016}%
  \BibitemOpen
  \bibfield  {author} {\bibinfo {author} {\bibfnamefont {I.~I.}\ \bibnamefont
  {Fishchuk}}, \bibinfo {author} {\bibfnamefont {A.}~\bibnamefont
  {Kadashchuk}}, \bibinfo {author} {\bibfnamefont {A.}~\bibnamefont
  {Bhoolokam}}, \bibinfo {author} {\bibfnamefont {A.}~\bibnamefont
  {de~Jamblinne~de Meux}}, \bibinfo {author} {\bibfnamefont {G.}~\bibnamefont
  {Pourtois}}, \bibinfo {author} {\bibfnamefont {M.~M.}\ \bibnamefont
  {Gavrilyuk}}, \bibinfo {author} {\bibfnamefont {A.}~\bibnamefont {K\"ohler}},
  \bibinfo {author} {\bibfnamefont {H.}~\bibnamefont {B\"assler}}, \bibinfo
  {author} {\bibfnamefont {P.}~\bibnamefont {Heremans}}, \ and\ \bibinfo
  {author} {\bibfnamefont {J.}~\bibnamefont {Genoe}},\ }\href {\doibase
  10.1103/PhysRevB.93.195204} {\bibfield  {journal} {\bibinfo  {journal} {Phys.
  Rev. B}\ }\textbf {\bibinfo {volume} {93}},\ \bibinfo {pages} {195204}
  (\bibinfo {year} {2016})}\BibitemShut {NoStop}%
\bibitem [{\citenamefont {Nenashev}\ \emph {et~al.}(2013)\citenamefont
  {Nenashev}, \citenamefont {Jansson}, \citenamefont {Oelerich}, \citenamefont
  {Huemmer}, \citenamefont {Dvurechenskii}, \citenamefont {Gebhard},\ and\
  \citenamefont {Baranovskii}}]{Nenashev2013}%
  \BibitemOpen
  \bibfield  {author} {\bibinfo {author} {\bibfnamefont {A.~V.}\ \bibnamefont
  {Nenashev}}, \bibinfo {author} {\bibfnamefont {F.}~\bibnamefont {Jansson}},
  \bibinfo {author} {\bibfnamefont {J.~O.}\ \bibnamefont {Oelerich}}, \bibinfo
  {author} {\bibfnamefont {D.}~\bibnamefont {Huemmer}}, \bibinfo {author}
  {\bibfnamefont {A.~V.}\ \bibnamefont {Dvurechenskii}}, \bibinfo {author}
  {\bibfnamefont {F.}~\bibnamefont {Gebhard}}, \ and\ \bibinfo {author}
  {\bibfnamefont {S.~D.}\ \bibnamefont {Baranovskii}},\ }\href@noop {}
  {\bibfield  {journal} {\bibinfo  {journal} {Phys. Rev. B}\ }\textbf {\bibinfo
  {volume} {87}},\ \bibinfo {pages} {235204} (\bibinfo {year}
  {2013})}\BibitemShut {NoStop}%
\bibitem [{\citenamefont {Wang}\ \emph {et~al.}(2013)\citenamefont {Wang},
  \citenamefont {Zhou}, \citenamefont {Zhang}, \citenamefont {Garoni},\ and\
  \citenamefont {Deng}}]{Wang2013}%
  \BibitemOpen
  \bibfield  {author} {\bibinfo {author} {\bibfnamefont {J.}~\bibnamefont
  {Wang}}, \bibinfo {author} {\bibfnamefont {Z.}~\bibnamefont {Zhou}}, \bibinfo
  {author} {\bibfnamefont {W.}~\bibnamefont {Zhang}}, \bibinfo {author}
  {\bibfnamefont {T.~M.}\ \bibnamefont {Garoni}}, \ and\ \bibinfo {author}
  {\bibfnamefont {Y.}~\bibnamefont {Deng}},\ }\href {\doibase
  10.1103/PhysRevE.87.052107} {\bibfield  {journal} {\bibinfo  {journal} {Phys.
  Rev. E}\ }\textbf {\bibinfo {volume} {87}},\ \bibinfo {pages} {052107}
  (\bibinfo {year} {2013})}\BibitemShut {NoStop}%
\bibitem [{\citenamefont {Hu}\ \emph {et~al.}(2014)\citenamefont {Hu},
  \citenamefont {Bl\"ote}, \citenamefont {Ziff},\ and\ \citenamefont
  {Deng}}]{Hao2014}%
  \BibitemOpen
  \bibfield  {author} {\bibinfo {author} {\bibfnamefont {H.}~\bibnamefont
  {Hu}}, \bibinfo {author} {\bibfnamefont {H.~W.~J.}\ \bibnamefont {Bl\"ote}},
  \bibinfo {author} {\bibfnamefont {R.~M.}\ \bibnamefont {Ziff}}, \ and\
  \bibinfo {author} {\bibfnamefont {Y.}~\bibnamefont {Deng}},\ }\href {\doibase
  10.1103/PhysRevE.90.042106} {\bibfield  {journal} {\bibinfo  {journal} {Phys.
  Rev. E}\ }\textbf {\bibinfo {volume} {90}},\ \bibinfo {pages} {042106}
  (\bibinfo {year} {2014})}\BibitemShut {NoStop}%
\bibitem [{\citenamefont {Koza}\ and\ \citenamefont
  {Po{\l}a}(2016)}]{Koza_2016}%
  \BibitemOpen
  \bibfield  {author} {\bibinfo {author} {\bibfnamefont {Z.}~\bibnamefont
  {Koza}}\ and\ \bibinfo {author} {\bibfnamefont {J.}~\bibnamefont {Po{\l}a}},\
  }\href {\doibase 10.1088/1742-5468/2016/10/103206} {\bibfield  {journal}
  {\bibinfo  {journal} {J. Stat. Mech.: Theory and Experiment}\ }\textbf
  {\bibinfo {volume} {2016}},\ \bibinfo {pages} {103206} (\bibinfo {year}
  {2016})}\BibitemShut {NoStop}%
\bibitem [{\citenamefont {Germs}\ \emph {et~al.}(2011)\citenamefont {Germs},
  \citenamefont {van~der Holst}, \citenamefont {van Mensfoort}, \citenamefont
  {Bobbert},\ and\ \citenamefont {Coehoorn}}]{Germs2011}%
  \BibitemOpen
  \bibfield  {author} {\bibinfo {author} {\bibfnamefont {W.~C.}\ \bibnamefont
  {Germs}}, \bibinfo {author} {\bibfnamefont {J.~J.~M.}\ \bibnamefont {van~der
  Holst}}, \bibinfo {author} {\bibfnamefont {S.~L.~M.}\ \bibnamefont {van
  Mensfoort}}, \bibinfo {author} {\bibfnamefont {P.~A.}\ \bibnamefont
  {Bobbert}}, \ and\ \bibinfo {author} {\bibfnamefont {R.}~\bibnamefont
  {Coehoorn}},\ }\href@noop {} {\bibfield  {journal} {\bibinfo  {journal}
  {Phys. Rev. B}\ }\textbf {\bibinfo {volume} {84}},\ \bibinfo {pages} {165210}
  (\bibinfo {year} {2011})}\BibitemShut {NoStop}%
\bibitem [{\citenamefont {Nakamura}(1984)}]{Nakamura1984}%
  \BibitemOpen
  \bibfield  {author} {\bibinfo {author} {\bibfnamefont {M.}~\bibnamefont
  {Nakamura}},\ }\href {\doibase 10.1103/PhysRevB.29.3691} {\bibfield
  {journal} {\bibinfo  {journal} {Phys. Rev. B}\ }\textbf {\bibinfo {volume}
  {29}},\ \bibinfo {pages} {3691} (\bibinfo {year} {1984})}\BibitemShut
  {NoStop}%
\bibitem [{\citenamefont {Chen}\ and\ \citenamefont {Schuh}(2006)}]{CHEN2006}%
  \BibitemOpen
  \bibfield  {author} {\bibinfo {author} {\bibfnamefont {Y.}~\bibnamefont
  {Chen}}\ and\ \bibinfo {author} {\bibfnamefont {C.~A.}\ \bibnamefont
  {Schuh}},\ }\href {\doibase https://doi.org/10.1016/j.actamat.2006.06.011}
  {\bibfield  {journal} {\bibinfo  {journal} {Acta Materialia}\ }\textbf
  {\bibinfo {volume} {54}},\ \bibinfo {pages} {4709} (\bibinfo {year}
  {2006})}\BibitemShut {NoStop}%
\end{thebibliography}%

\end{document}